\documentclass{article}
\usepackage{spconf,amsmath,graphicx}

\usepackage{booktabs}

\usepackage{enumitem}
\setlist{nosep, leftmargin=14pt}

\usepackage{mwe} % to get dummy images

\title{Gland Segmentation using SAM with Cancer Grade as a Prompt}
\name{{Yijie Zhu$^{\star}$, Shan E Ahmed Raza$^{\star}$}}
\address{$^{\star}$Tissue Image Analytics Centre, Department of Computer Science, University of Warwick}
\begin{document}
%\ninept
%
\maketitle
\begin{abstract}
Cancer grade is a critical clinical criterion that can be used to determine the degree of cancer malignancy. Revealing the condition of the glands, a precise gland segmentation can assist in a more effective cancer grade classification. In machine learning, binary classification information about glands (i.e., benign and malignant) can be utilized as a prompt for gland segmentation and cancer grade classification. By incorporating prior knowledge of the benign or malignant classification of the gland, the model can anticipate the likely appearance of the target, leading to better segmentation performance. We utilize Segment Anything Model to solve the segmentation task, by taking advantage of its prompt function and applying appropriate modifications to the model structure and training strategies. We improve the results from fine-tuned Segment Anything Model and produce SOTA results using this approach.
\end{abstract}
\section{Introduction}
\label{sec:intro}

Colorectal cancer is one of the most common cancer types, with a notably high mortality rate. A large proportion of colorectal cancers are classified as adenocarcinomas \cite{alzahrani2021general}. Colorectal adenocarcinoma is distinguished by glandular formation. Pathologists rely on the morphology of glands and the degree of glandular formation as factors to evaluate the degree of tumor differentiation and classify cancer grade \cite{fleming2012colorectal}. The segmentation of glands plays a significant role in automated and objective grading of cancer. Accurate segmentation aids in deriving quantitative measurements, and a more comprehensive evaluation of the tumor’s malignancy can be achieved, leading to better diagnostic accuracy. Several models are proposed to address the challenge of gland segmentation. Graham et al. \cite{graham2019mild} proposed MILD-Net. Original images are resized and concatenated with feature maps in different stages of the model to minimize information loss. Wen et al. \cite{9164951} proposed GCSBA-Net to capture more comprehensive feature from images by using a Gobar-based encoder to extract texture information and applying the Bi-Attention mechanism to learn spatial and channel information.

As a foundation model, Segment Anything Model (SAM) \cite{Kirillov_2023_ICCV} is dedicated to the tasks of image segmentation. Primarily trained on a broad dataset, it demonstrates impressive zero-shot performance involving natural images. However, when applied to more specialized domains, such as medical imaging, this zero-shot performance tends to decline \cite{deng2023segment}. Despite this, the strong generalization ability of SAM provides a promising foundation for adaption \cite{ma2024segment}. By fine-tuning or introducing task-specific structural modifications, SAM can be tailored to address a range of downstream segmentation tasks in medical imaging \cite{wu2023medical}. Moreover, SAM is designed to incorporate prompts. The model can be guided to focus on the specific area of interest by providing prompts, which in turn can enhance its performance \cite{mazurowski2023segment}.

The main challenge of gland segmentation is the wide-ranging variations in the morphological features of glands of different grades \cite{awan2017glandular}. However, when the task is limited to binary classification, distinguishing between benign and malignant cases, these variations can instead assist the classification model in making accurate predictions. By incorporating previously predicted benignity or malignancy of the gland as prior knowledge into the segmentation model, the model can make an initial informed judgment, thereby improving its performance. CGS-Net proposed by Tang et al. \cite{tang2023cgs} utilizes the segmentation encoder to simultaneously perform benign and malignant classification. Then each layer of the encoder obtains feature maps enriched with classification information, which are passed to the corresponding decoder layers.

The paper has following contributions: 1) SAMs of different scales are fine-tuned to assess their performance on gland segmentation tasks. 2) A new method to provide grade prompt is designed. The heat map generated from the classification model using Grad-CAM++ (which contains benign and malignant information) is provided to the segmentation model which improves segmentation performance. 3) A prompt adapter is developed to process the prompt. 4) Structure modifications and adjustments to training strategies are applied to the SAM, enabling the proposed model to better adapt to the gland segmentation task.

\section{Methods}
\label{sec:methods}

The proposed model shown in Fig. 1 is designed to generate predictions for both cancer classification and gland segmentation simultaneously through two branches.

\begin{figure}[ht]
    \centering
    \includegraphics[width=8.7cm,height=7cm]{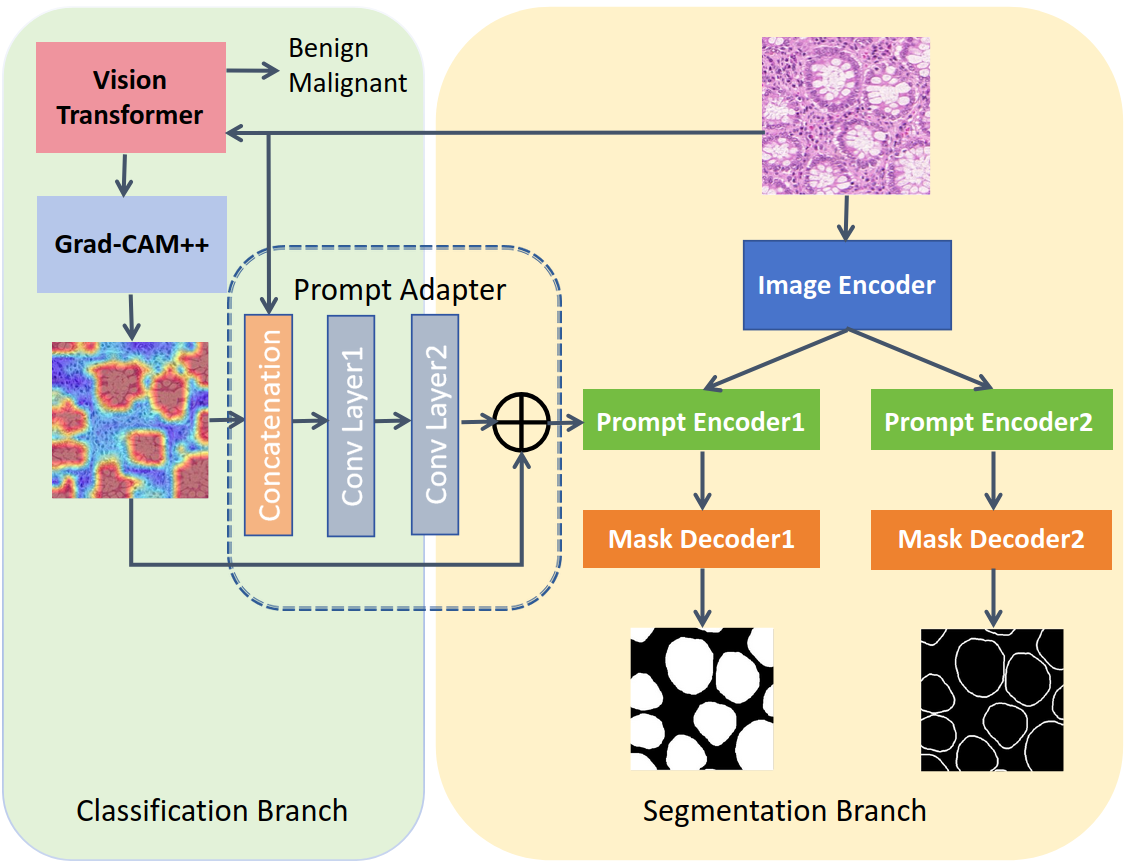}  
    \caption{The classification branch determines whether the gland is benign or malignant and generates a heat map. The prompt adapter processes the heat map from the classification branch and sends it to the segmentation branch. The segmentation branch focuses on distinguishing the gland and contour areas from the background area.}
    \label{fig:your-label}
\end{figure}

\subsection{Classification Branch}

Within the classification branch, the benign or malignant status of glands is predicted. Vision Transformer (ViT) \cite{dosovitskiy2021imageworth16x16words} is utilized to complete the classification task. ViT model is pre-trained on large datasets, which offers the possibility of transfer learning. It can be fine-tuned on specific pathological datasets.

\subsection{Prompt of Cancer Grade}

In the proposed model, a new method to provide grade prompt is utilized. When the segmentation model is informed about the specific type of gland in advance, it can pay attention to the characteristics of such gland category. The prompt allows the model to optimize its ability in gland segmentation.

The prior knowledge can be sourced from the classification branch of the model. When the ViT model classifies an image, particular regions that the model deems critical are focused on. Details and patterns that can indicate benignity or malignancy of the glands are largely contained in the regions.

Grad-CAM++ \cite{chattopadhay2018grad} is a technique that can explain classification prediction of ViT by visualizing the contribution of every part of an image to the result. The heat map provided by Grad-CAM++ sets a value for each image pixel representing the region's significance. As shown in Fig. 2, the gland areas are mostly considered important by ViT in classification. Given that the Grad-CAM++ heat map is a one-channel matrix and maintains the same size as the original input image, it can be used as a prompt. 

\begin{figure}[t]
    \centering
    \includegraphics[width=6.5cm,height=4.5cm]{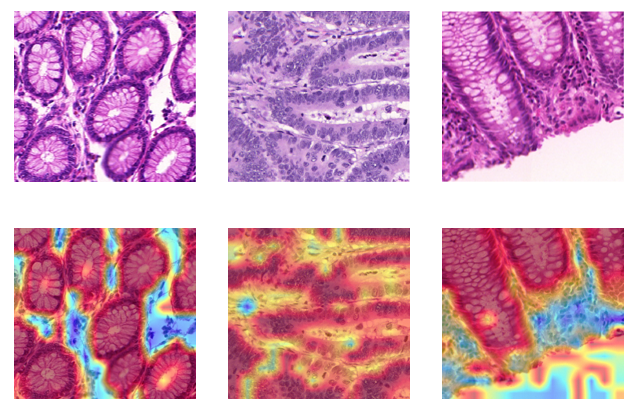}  
    \caption{The heat maps of the images that generated from classification branch with Grad-CAM++ are shown.}
    \label{fig:your-label}
\end{figure}

% The values in the heat map matrix represent the probability that the corresponding pixel in the image belongs to a gland. 

The mask is then relayed to the segmentation branch. The process can bridge the two branches and enhance the information available for the segmentation branch.

\subsection{Prompt Adapter}

Before the input is passed into the segmentation branch, the heat map is first processed by a designed adapter. This adapter facilitates the effective integration of the information from the heat map into the network. The heat map is first concatenated with the original image in the adapter. This process helps to provide detailed information from the original image corresponding to the heat map. The four-channel feature is then reduced to one channel through two convolutional layers, making the resulting prompt suitable for the network requirements. Batch normalization and RELU function are applied after the two convolutional layers. The heat map is added with the output from the second convolution to feed to the segmentation branch.

\subsection{Segmentation Branch}

SAM is selected as the foundation structure for the segmentation branch and is modified to output both gland and contour predictions. A common challenge in gland segmentation is that the neighboring glands may be predicted to be connected. An accurate contour prediction can address the challenge by removing the overlapping parts of the glands and the contours from the predicted glands.

A single image encoder is used in the branch and is responsible for extracting features from the input image. The model is then divided into a gland prediction branch and a contour prediction branch, each with its own prompt encoder and mask decoder. 

For contour prediction, no prompt is input into the prompt encoder. For gland prediction, the heat map generated from the classification branch via Grad-CAM++ and processed by the prompt adapter is used as the prompt. The prompt helps ensure that the prediction can more thoroughly cover the regions where glands are present.

Finally, the outputs from the two prompt encoders are separately fed into their respective mask decoders to predict the gland and contour simultaneously.

\subsection{Training Process}
 Transfer learning is applied to both classification and segmentation branches to provide a correct path for subsequent training.

For the classification branch, the ViT model is first fine-tuned to fit the cancer classification task before being used in the proposed model. The model achieved accuracy rates of 97.1\% and 98.7\% on the test set A and B.

For the segmentation branch, the parameters of a fine-tuned SAM, including the image encoder, prompt encoder and mask decoder of gland segmentation branch, are loaded into the proposed model. The parameters of SAM are loaded into the prompt encoder and mask decoder of contour segmentation branch.

 The shared image encoder, along with the prompt encoder, prompt adapter, and mask decoder of the gland segmentation branch, is first trained. To avoid influencing the previous training results of gland segmentation, the image encoder is frozen in the further training. At last, the prompt encoder and the mask decoder of the contour segmentation branch are trained.

\subsection{Loss Function}

When fine-tuning SAM, mean square error (MSE) loss was used. The concept of weight map, originally introduced in U-Net \cite{ronneberger2015unetconvolutionalnetworksbiomedical}, is applied to the gland annotation when training both gland and contour branch of proposed model. The weight map of gland annotation is multiplied element-wise with MSE loss for each pixel. The result is summed to obtain the weighted MSE loss.

\subsection{Post Processing}

The predicted pixel values of the overlapping area from different patches are averaged to obtain a unified value, which can help achieve smooth transitions between patches. A sigmoid layer outputs the prediction result of the model. A threshold of 0.5 is set to produce a binary mask prediction.

As is shown in Fig. 3, the overlapping regions between the contours and glands are first identified and then eliminated.

\begin{figure}[ht]
    \centering
    \includegraphics[width=8.5cm,height=2.5cm]{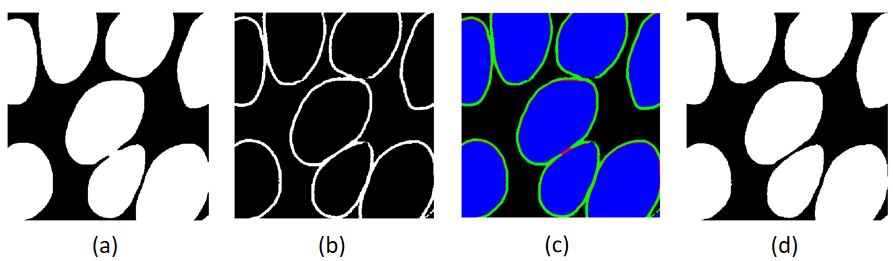}  
    \caption{(a) Gland prediction. (b) Contour prediction. (c) Gland prediction is shown in blue. Contour prediction is shown in green. The overlapping area is shown in red. (d) Result after removing the overlapping area.}
    \label{fig:your-label}
\end{figure}

A median filter is applied to the image to smooth the boundary. Small irregular dots in the background and minor holes in the foreground are considered noise and are removed.

\section{Experiment and Results}
\label{sec:experiment}

\subsection{Data Preparation}

The open-source Gland Segmentation Challenge (GlaS) \cite{sirinukunwattana2017gland} dataset was used in this study. There are 85 images in the training set. There are 60 and 20 images in test sets A and B. A square with a size of 400$\times$400 pixels was set as the input for the model. The images in GlaS dataset are all larger than this standard. Four partially overlapping images can be extracted from the four corners of one original image. Additionally, the extracted images are rotated by 90 degrees in varying multiples to increase the diversity of the data set and improve the robustness of the model during training. Contour annotations are produced by removing the eroded gland annotation from the dilated gland annotation.

\subsection{Model Settings}

For the classification branch, a pre-trained ViT is fine-tuned. The open-source ‘deit-base-patch16-224’ \cite{touvron2021training} is utilized. For the segmentation branch, pre-trained SAMs \cite{Kirillov_2023_ICCV} are provided at three different scales, which are ViT-B SAM (SAM-B), ViT-L SAM (SAM-L), and ViT-H SAM (SAM-H).

\begin{figure*}[ht]
    \centering
    \includegraphics[width=16cm,height=5.5cm]{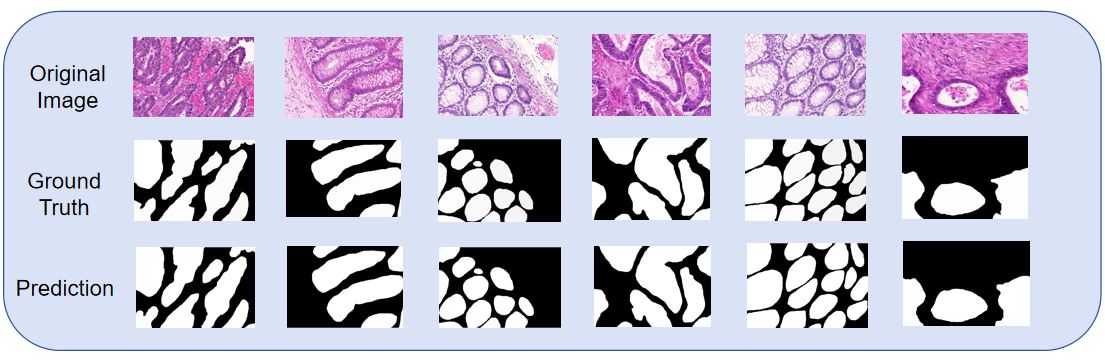}  
    \caption{Original images, ground truths, and predicted results are presented in the first, second, and third rows.}
    \label{fig:your-label}
\end{figure*}

\subsection{Evaluation Metrics}

Three metrics from the GlaS Challenge, including F1 Score, Object-level Dice, and Object-level Hausdorff, are used to assess the models.

Table 1 presents the metrics of the three fine-tuned models at different scales. As the model scale increases, the task performance improves, with the larger versions demonstrating superior results. 

\begin{table}[h]
\centering
\setlength{\tabcolsep}{3.5pt}
\caption{Comparison of Metrics with Fine-tuned SAM.}
\label{tab:comparison}
\renewcommand{\arraystretch}{1.3}
\fontsize{8}{12}\selectfont
\begin{tabular}{lcccccc}
\toprule
\textbf{Method} & \multicolumn{2}{c}{\textbf{F1 Score} $\boldsymbol{\uparrow}$ } & \multicolumn{2}{c}{\textbf{Object Dice} $\boldsymbol{\uparrow}$ } & \multicolumn{2}{c}{\textbf{Object Hausdorff} $\boldsymbol{\downarrow}$ } \\
 \cmidrule(r){2-3} \cmidrule(r){4-5} \cmidrule(r){6-7}
 & Test A & Test B & Test A & Test B & Test A & Test B \\
\midrule
\textbf{SAM-B}  & 0.880 & 0.764 & 0.884 & 0.813 & 61.384 & 121.047 \\
\textbf{Prompted SAM-B}  & \textbf{0.882} & \textbf{0.777} & \textbf{0.890} & \textbf{0.827} & \textbf{58.464} & \textbf{114.469} \\
\hline
\textbf{SAM-L}  & 0.925 & 0.810 & 0.914 & 0.846 & 43.380 & 103.227 \\
\textbf{Prompted SAM-L}  & \textbf{0.927} & \textbf{0.813} & \textbf{0.919} & 0.846 & \textbf{37.052} & \textbf{98.605} \\
\hline
\textbf{SAM-H}  & \textbf{0.932} & 0.820 & 0.917 & 0.879 & 42.441 & 77.158 \\
\textbf{Prompted SAM-H} & 0.929 & \textbf{0.841} & \textbf{0.921} & \textbf{0.881} & \textbf{41.189} & \textbf{74.300} \\
\bottomrule
\end{tabular}
\end{table}

In addition, the grade prompt is used on these three scales of the SAM. It is clear that the model of each scale exhibits improved performance through such modifications.

\begin{table}[h]
\centering
\setlength{\tabcolsep}{3.5pt}
\caption{Comparison of Metrics with GlaS \cite{sirinukunwattana2017gland} Benchmarks.}
\label{tab:comparison}
\renewcommand{\arraystretch}{1.3}
\fontsize{8}{12}\selectfont
\begin{tabular}{lcccccc}
\toprule
\textbf{Method} & \multicolumn{2}{c}{\textbf{F1 Score} $\boldsymbol{\uparrow}$} & \multicolumn{2}{c}{\textbf{Object Dice} $\boldsymbol{\uparrow}$} & \multicolumn{2}{c}{\textbf{Object Hausdorff} $\boldsymbol{\downarrow}$} \\
 \cmidrule(r){2-3} \cmidrule(r){4-5} \cmidrule(r){6-7}
 & Test A & Test B & Test A & Test B & Test A & Test B \\
\midrule

\textbf{Prompted SAM-H } & \textbf{0.929} & \textbf{0.841} & \textbf{0.921} & \textbf{0.881} & \textbf{41.189} & \textbf{74.300} \\

\textbf{CUMedVision2}  & 0.912 & 0.716 & 0.897 & 0.781 & 45.418 & 160.347 \\

\textbf{ExB1} & 0.891 & 0.703 & 0.882 & 0.786 & 57.413 & 145.575 \\
\textbf{ExB3} & 0.896 & 0.719 & 0.886 & 0.765 & 57.350 & 159.873 \\

\textbf{Freiburg2}  & 0.870 & 0.695 & 0.876 & 0.786 & 57.093 & 148.463 \\

\textbf{CUMedVision1}  & 0.868 & 0.769& 0.867 & 0.800 & 74.596 & 153.646 \\

\textbf{ExB2}  & 0.892 & 0.686 & 0.884 & 0.754 & 54.785 & 187.442 \\

\textbf{Freiburg1}  & 0.834 & 0.605 & 0.875 & 0.783 & 57.194 & 146.607 \\

\bottomrule
\end{tabular}
\end{table}
The prompted SAM-H is compared with the benchmarks listed in GlaS contest in Table 2, achieving the best performance across each metrics.

\subsection{Results on Whole Slide Images}

Fig. 4 shows the prompted SAM-H prediction results on the GlaS dataset's test sets. It reveals that the proposed model performed well on the test sets, and there is a substantial similarity between the model prediction and the provided ground truth. Glands with both regular and irregular shapes are accurately predicted, and in cases where adjacent glands are very close to each other, they are successfully separated.

Furthermore, columns 2, 3 and 5 represent benign cancer samples, while columns 1, 4, and 6 are malignant samples. The model exhibited consistent and stable performance in gland segmentation tasks, handling benign and malignant cancer cases with high precision.

\section{Conclusion}

As benign and malignant glands exhibit varying morphological characteristics, a deep learning model can take advantage of cancer grade information as prior knowledge. A new method to provide grade prompt is designed. Heat maps highlighting regions generated by Grad-CAM++ serve as a prompt to guide the segmentation model. A prompt adapter is proposed to help model adapt to the heat map prompt. By increasing the number of prompt encoders and mask decoders to two, the model can make predictions for gland and contour segmentation at the same time. Removing the overlapping area between glands and contours can help separate adjacent glands. A stepwise training approach enables the model to achieve improved training outcomes.

In future, we plan to test the performance with different loss functions, such as Cross Entropy, Dice and AC Loss. Besides, Hi-gMISnet \cite{showrav2024hi} achieved an F1 score of 0.932 on the overall test set, demonstrating the potential of generative models, which can be further explored. Beyond gland segmentation, the purposed method can be applied to broader tasks, such as cell segmentation, with experiments conducted on datasets like PanNuke \cite{gamper2019pannuke}.

As a result, the proposed model successfully and effectively addressed the task of gland classification and segmentation. Compared with the fine-tuned Segment Anything Model of different scales, the modified model performs better.

\section{Compliance with ethical standards}

The research was conducted using publicly available GlaS Challenge data set \cite{sirinukunwattana2017gland}.

\section{Acknowledgment}

Yijie Zhu is funded by China Scholarship Council - University of Warwick Scholarship.
\bibliographystyle{IEEEbib}
\bibliography{strings,refs}

\end{document}